\documentclass{elsart}
\usepackage{graphicx}
\journal{Physica D}

\begin{document}

\begin{frontmatter}

\title{Discriminating dynamical from additive noise in the 
Van der Pol oscillator}

\author[Alicante]{C. Degli Esposti Boschi},
%\ead{esposti@ua.es}
\author[Quilmes]{G.J. Ortega} and
\author[Alicante]{E. Louis}

\address[Alicante]{Departamento de F\'{\i}sica Aplicada and Unidad Asociada
of the Consejo Superior de Investigaciones Cient\'{\i}ficas, \\
Facultad de Ciencias, Universidad de Alicante, Apartado 99, E-03080,
Alicante, Spain}

\address[Quilmes]{Centro de Estudios e Investigaciones,
Universidad Nacional de Quilmes,\\ R. S. Pe\~na 180, 1876, Bernal, Argentine}

\date{\today}

%\maketitle

\begin{abstract}
We address the distinction between dynamical and additive noise
in time series analysis by making a joint evaluation of both 
the statistical continuity of the series and the statistical
differentiability 
of the reconstructed measure. Low levels of the latter and
high levels of the former indicate the presence
of dynamical noise only, while low values of the two are
observed as soon as additive noise contaminates the signal.
The method is
presented through the example of the
Van der Pol oscillator, but is expected to be of general
validity for continuous-time systems.
\end{abstract}
%\vspace*{2.0truecm}

\end{frontmatter}

\section{Introduction}
\label{S_intro}
Experimental time series are always blurred with additive noise coming 
from a variety of sources \cite{WG94,KS97}. Additive noise further complicates 
time series analysis, leading to ambiguous interpretations
of the basic quantities which characterise the dynamics, like
correlation dimension and Lyapunov exponents
\cite{KANTZ,SY99}.
In particular it may lead unreachable the goal of 
differentiating deterministic 
from stochastic dynamics \cite{SM90,KG92,WB93,SC94,BK99,OL98,OL00}.

\par Here an attempt is made to discriminate additive noise (AN) 
from the dynamical
noise (DN) the system may have. Note that the term {\it additive}
used here denotes a component of noise that is superimposed to
the underlying ``clean'' signal, due to the measurement process.
Common synonyms found in the literature are {\it measurement}
or {\it observational} noise.
It should not be confused with the term used in the context
of stochastic processes (see, for istance, \cite{TSM}) where
it often indicates a coordinate-independent stochastic forcing
as opposite to {\it multiplicative} noise, for which the amplitude
of the random terms depends on the coordinates themselves.
Here we do not focus on this last distinction, and we will denote
generically as {\it dynamical} every noisy mechanism intrinsic to
the system.

Our method is based upon a fundamental,
topological, property of deterministic systems, namely, the differentiability
of the measure along the reconstructed trajectory \cite{PC95,OL98,OL00} or,
more specifically, the continuity of its logarithmic derivative. 
Starting from this 
basis we show that noise (additive or dynamical) decrease the
differentiability of the measure. This would in principle hinder the
possibility of differentiating the two types of noise. Then we look
at the same property (continuity) of the coordinate. Now, while
AN destroys continuity of the coordinate, the latter
is scarcely affected by DN. Thus, a low continuity of
both the coordinate and (the log derivative of) 
the reconstructed measure would indicate
the presence of AN, while if continuity is high
for the coordinate and low for the measure the system contains DN only
(see following section). The method, however, doesn't differentiate 
sharply a system with the
two types of noise from one having solely AN. 
Notwithstanding,
we believe that the present approach is a significant step forward in
the understanding of determinism and stochasticity.
To illustrate how the method works we consider the simple case
of the Van der Pol oscillator \cite{SY97}. No reasons are forseen that may
indicate that the applicability of the method is limited to simple,
non chaotic, dynamical systems, such as the one investigated here.

The rest of the paper is organised as follows. In section \ref{S_Gc} 
we make 
some general considerations on the effects of additive and dynamical noise.
In section \ref{S_MNP} we briefly discuss the numerical methods followed
to solve the dynamical equations with noise and the embedding process.
The procedure followed to evaluate the
(statistical) continuity
is also summarised in that section. The results are discussed 
in section \ref{S_R}, while section \ref{S_Conc} 
is devoted to the conclusions of our work. 

\section{General considerations}
\label{S_Gc}
The method of the reconstructed measure
along the trajectory \cite{OL98,OL00} has the capability of revealing
the presence of DN in a given time series by
looking at the degree of (statistical) continuity
of the logarithmic derivative of such a measure. 
To this end, it is to be considered as complementary
to other techniques, essentially based on
short-time predictability \cite{SM90} or
on smoothness in phase-space \cite{KG92,WB93,SC94}.
As far as AN is concerned, in ref. \cite{Mi96} it is
argued that methods of the second type can be useful even
in real experimental series affected by measurement components.
A different line is that followed by Barahona and Poon \cite{BP96},
who are able to cope with rather large amounts of AN, at the ``price''
of building the analysis upon a given class of Volterra-type nonlinear
models adjusted on the time series itself.
Here, we do not specify any {\it a priori} dynamics and
pursue the extension of the method of Ortega and Louis
\cite{OL98,OL00} to deal with AN. The choice is motivated
by very recent results \cite{HDIM} which allow us to
believe that this method is suitable even for high-dimensional
chaotic systems, which somehow fool the above-mentioned
alternative approaches.

As in refs. \cite{OL98,OL00} we concentrate
on continuous-time systems and start by observing that 
a DN-term modeled by (the increments of)
a Wiener process $\delta w_k$ (coupled through
some constants $G_{jk}$):
\begin{equation}
{\rm d}x_j=F_j(\vec{x}) {\rm d}t+\sum_{k} G_{jk} \delta w_k
\label{SDEs}
\end{equation}
is basically different from white AN, 
superimposed to the time series $\{ s_n=s[\vec{x}(n \Delta t)] \}$ after
a clean ($G_{jk}=0$) integration:
\begin{equation}
\bar{s}_n = s_n+\eta_n \; .
\label{ansn}
\end{equation}
This is seen in the typical case in which
the ``measurement function'' $s(\vec{x})$
is simply one of the coordinates, say $x_j$.
In fact, while the white-noise process $\eta_n$
of eq. (\ref{ansn}) appears to be always
{\it discontinuous} in time, the $\delta w_k$ in eq. (\ref{SDEs})
lead to a {\it continuous} solution, as it can be
seen from the increments in a time $\delta t$:
\begin{equation}
x_j(t+\delta t)=x_j(t)+F_j \delta t+\sum_k G_{jk} \gamma_k 
(\delta t)^{1/2} + \cdots
\label{ixj}
\end{equation}
The $\gamma_k$'s are $\delta$-correlated normal Gaussian random numbers,
and in the limit $\delta t \rightarrow 0$
one has $x_j(t+\delta t) \rightarrow x_j(t)$ \cite{NOTE}.
The same applies for a generic (at least differentiable)
measurement function, as can be seen from a first-order expansion:
\begin{equation}
s[\vec{x}(t+\delta t)]=s[\vec{x}(t)]+\sum_j
(\partial_j s)
[F_j \delta t + \sum_k G_{jk} \gamma_k {(\delta t)}^{1/2}] \; .
\label{imfs}
\end{equation}

Apart from this basic difference, the distinction
between AN and DN
in the real output of a numerical integration (or
of an experimental device) is still a complicated task
because the continuity must be judged on the basis
of {\it finite} increments $\Delta t$ over which
the series acquires a finite increment $\Delta s$.
From this point of view, the continuity statistics (CS) of Pecora
{\it et al.} \cite{OL00,PC95} sheds some light as it evaluates the degree
of continuity at different resolution scales.

Before closing the section we draw some comments on another
type of problem which is sometimes put forth \cite{Sz93,KS97}, namely,
that the distinction between AN and DN is not well-posed
since (at least in some cases) they can be mapped onto each other.
Consider a discrete-time dynamics with some noisy feedback $\vec{\gamma}$:
\begin{equation}
\vec{x}_{n+1}=\vec{f}\left(\vec{x}_n\right)+\vec{\gamma}_{n+1} \; , 
\; \; n=0,1,\dots
\label{n_mapdyn}
\end{equation}
In absence of DN the clean time series
from the $j$-th coordinate would be:
\begin{equation}
s_0=x_{0j} \, , \; s_1=f_j\left(\vec{x}_0\right) , \;
s_2=f_j\left( \vec{f}\left(
\vec{x}_0 \right) \right) , \;
s_3=f_j\left( \vec{f}\left( \vec{f}\left(\vec{x}_0
\right) \right) \right) , \; \dots
\label{cmts}
\end{equation}
whereas DN turns it to:
\begin{displaymath}
\bar{s}_0=x_{0j} \, , \; 
\bar{s}_1=f_j\left(\vec{x}_0\right) + \gamma_{1j} , \;
\bar{s}_2=f_j\left(\vec{f}\left(\vec{x}_0\right)+\vec{\gamma}_1\right)+
\gamma_{2j} ,
\end{displaymath}
\begin{equation}
\bar{s}_3=f_j\left( \vec{f}\left( 
\vec{f}\left(\vec{x}_0\right)+\vec{\gamma}_1 \right)+\vec{\gamma}_2 \right)+\gamma_{3j} \; \dots
\label{nmts}
\end{equation}
Now, by the light of eq. (\ref{ansn}), one could equally say that
the time series (\ref{nmts}) stems from a deterministic dynamics
$\vec{x}_{n+1}=\vec{f}\left( \vec{x}_n \right)$ plus
the following (equivalent) AN:
\begin{displaymath}
\eta_1=\gamma_{1j} , \; \eta_2=\gamma_{2j}+f_j\left(\vec{f}\left(\vec{x}_0\right)+\vec{\gamma}_1\right)-f_j\left( \vec{f}\left(
\vec{x}_0 \right) \right) , \;
\end{displaymath}
\begin{equation}
\eta_3=\gamma_{3j}+f_j\left( \vec{f}\left( 
\vec{f}\left(\vec{x}_0\right)+\vec{\gamma}_1 \right)+\vec{\gamma}_2 \right)-
f_j\left( \vec{f}\left( \vec{f}\left(\vec{x}_0
\right) \right) \right)
\label{mandnm}
\end{equation}
Similar arguments apply to more general measurement functions
and also to continuous-time systems, at least at the leading order of
eq. (\ref{imfs}). Thus, in principle it is conceivable that
a special type of AN can mimic a given DN present in the dynamics.
However, from a practical point of view, a measurement noise
like the one in eq. (\ref{mandnm}) is not too realistic, as
it is ``customised'' on the dynamics itself (through $\vec{f}$)
and on the initial conditions (in the case of continuous-time systems
it would also scale with the integration time).
In addition, what is more important is that the process
in eq. (\ref{mandnm}) is autocorrelated, as can be seen
by inverting the relationship between the $\vec{\gamma}$'s and
the $\eta$'s step by step. Again, the autocorrelation pattern
is determined by dynamical features of the system in a complicated
way and in general it does not correspond to a typical
measurement component. Thus, being conscious of the mathematical
subtleties which may arise in rigorous framework,
in what follows we restrict to
white noise processes as a first possible modelisation of
real situations.

\section{Model and Numerical Procedures}
\label{S_MNP}

\subsection{The  Van der Pol oscillator}
\label{S_VdPo}
The different effects of AN and DN have been
tested on the Van der Pol system, whose
``clean'' equations are:
\begin{displaymath}
\dot{y}=-\mu (x^2-1)y-x
\end{displaymath}
\begin{equation}
\dot{x}=y
\label{VdPsystem}
\end{equation}
The white noise, added either to the left-hand sides of
eqs. (\ref{VdPsystem}) or to the clean coordinate
$x$, is generated using a Gaussian distribution
with zero mean. Different values of the variance,
$\sigma^2$, have been considered in order to tune
the strength of the noisy terms. 
We point out that from the discussion of the previous section,
and from the following results, there are no apparent obstacles
to extend these ideas to high-dimensional/chaotic systems.
The reason why we choose such a simple system is that, due
to its one-dimensional attractor, we can clarify
the points without worrying about the problems of
``wandering'' \cite{OL00} which may arise when the
embedding dimension is pushed at higher and higher values.

\subsection{Numerics}
\label{S_Num}
The initial conditions for the system (\ref{VdPsystem}) have been
set to $(x_0=1,y_0=0)$ and $\mu=1$. These choices
lead to a ``clean'' amplitude of oscillation
of about 2, with period 6.6.
Dealing with stochastic differential equations, we
used an Euler integration scheme to produce time-series 
of 16384 points from the $x$
coordinate. To achieve a good compromise
in terms of numerical accuracy we choosed an 
integration time step $\Delta t = 0.01$.
The transient time was observed to
be sufficient for the oscillator to relax on
the 1D attractor. Numerically we explored embedding dimension $m$
up to 10, using a delay time of 166 which
corresponds to the first zero of the autocorrelation
function (in units of $\Delta t$).
As in \cite{OL00}, we adopt an Epanechnikov kernel measure
estimator with a radius of 0.05 (after that the
reconstructed attractor has been rescaled to lie
within the hypercube $[0,1] \times \cdots \times [0,1]$). 

\subsection{Continuity statistics}
\label{S_CS}
In order to test the mathematical properties embodied in
a possible mapping between two given time series, that is, 
continuity, differentiability, inverse
differentiability and injectivity, Pecora {\it et al.} \cite{PC95} have
developed a set of statistics aimed to test quantitatively
these features. Their algorithms are of general use and can
in particular be applied to test topological
properties in any pair of sets of points.
Basically, the method is intended to evaluate,
in  terms of probability or confidence levels, whether two data
sets are related by a mapping having the continuity property:
A function $f$ is said to be
continuous at a point $\vec{x}_0$ if $\forall \varepsilon > 0, \exists
\delta > 0$ such that $\parallel \vec{x} - \vec{x}_0 \parallel <
\delta \Rightarrow \parallel f(\vec{x})) - f(\vec{x}_0) \parallel
< \varepsilon$. The results are tested against the null--hypothesis,
specifically, the case in which no functional relation exists.
This is done by means of the statistics proposed by Pecora {\it et al.}
\cite{PC95}
\begin{eqnarray}
\Theta_{C^0} (\varepsilon) = \frac{1}{n_p} \sum_{j=1}^{n_p}
\Theta_{C^0} (\varepsilon , j)
\label{e7}
\end{eqnarray}
and
\begin{eqnarray}
\Theta_{C^0} (\varepsilon, j) = 1 - \frac{p_j}{P_{\rm max}}
\label{e8}
\end{eqnarray}
\noindent
where $p_j$ is the probability that all of $n_\delta$ points in the
$\delta$-set, around a certain point $\vec{x}_j$, fall in the
$\varepsilon$-set around $f(\vec{x}_j)$.
The likelihood that this will happen must
be relative to the probability, $P_{\rm max}$, of the
most likely event under the null hypothesis.
In the Appendix we present some details on the calculation
of the ratio in eq. (\ref{e8}), useful for a numerical implementation.
The sum in
eq. (\ref{e7}) represent an average over $n_p$ points
chosen at random in the whole time series. 
Now, when $\Theta_{C^0} (\varepsilon)
\approx 1$ we can confidently reject the null hypothesis, and
assume that there exists a continuous function.
As in the work of Pecora {\it et al.} \cite{PC95} the $\varepsilon$
scale is relative to the standard deviation of the time series,
and thus $\varepsilon \in [0,1]$.
Plots of $\Theta_{C^0}(\varepsilon)$ versus $\varepsilon$ can be used to
quantify the degree of statistical continuity of a given function.
In order to characterise the continuity statistics by means of
a single parameter we have also calculated,
\begin{equation}
\theta = \int_0^1 \Theta_{C^0} (\varepsilon) {\rm d}\varepsilon
\end{equation}
\noindent The limiting values of $\theta$, namely, 0 and 1, correspond
to a strongly discontinuous and a fully continuous function, respectively.
In the results reported here the function $f$ mentioned above can
be either the logarithmic derivative of the measure
or the coordinate itself. When useful, we will denote the corresponding
continuity statistics by CSLDM and CSC.
The resolution scale $\varepsilon$ varied in the
range $[0.0001,1]$ and
$n_p$ was always fixed to be a 20\% of the total length of the series.

\section{Results}
\label{S_R}
\begin{figure}
\begin{center}
\includegraphics*[width=8cm,height=12cm,angle=270]{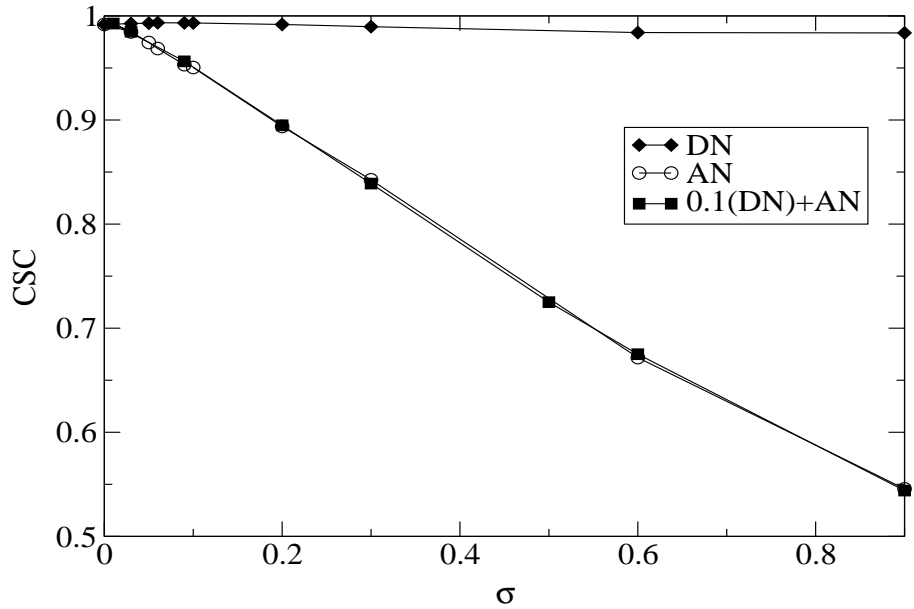}
%\
%\psfig{file=CS_x-vs-noise.eps,height=2.in}
\caption{CSC versus noise standard deviation, 
$\sigma$, for the $x$ coordinate of the Van der Pol oscillator. 
Results are shown for pure AN (no DN, open circles), pure 
DN (no AN, full diamonds) and for a system having a 
fixed level $\sigma_{\rm DN}=0.1$ plus AN
with variable standard deviation (full squares, almost overlapped 
with the open circles).}
\label{CSCvs}
\end{center}
\end{figure}

\begin{figure}
\begin{center}
\includegraphics*[width=8cm,height=12cm,angle=270]{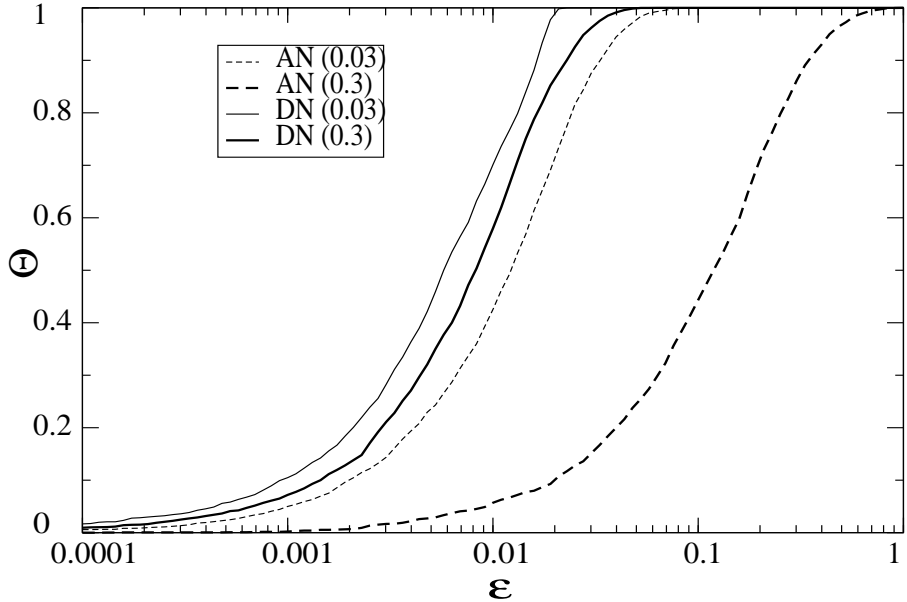}
\caption{CSC resolved at different $\varepsilon$ scales. The legend
indicates the type and the levels (standard deviation) of noise.}
\label{CSCve}
\end{center}
\end{figure}

The CSC for the amplitude $x$ of the Van der Pol oscillator
(eqs. (\ref{VdPsystem})) is shown in fig. \ref{CSCvs} 
for different levels and kinds of noise.
On the one hand it is readily seen that the CSC for
the time series affected by DN is essentially independent
of the noise amplitude. This appears to be consistent with
the continuity analysis sketched in section \ref{S_Gc}.
On the other hand, the CSC for series affected by
AN decreases steadily, and almost linearly, 
with the noise level $\sigma_{\rm AN}$.
We have considered also the more subtle case in which
both AN and DN are present. In particular, we have fixed
a moderate level of DN, like $\sigma_{\rm DN}=0.1$, and
then contaminated the resulting $x$-time series with various
levels of AN. The corresponding points in fig. \ref{CSCvs}
are almost indistinguishable from those without
DN, indicating that what rules the CSC is the presence
of AN. To give an idea of how the CS is disributed over
the different resolution scales we have plotted, in fig. \ref{CSCve}, 
the $\Theta_{C^0}$ statistics of the coordinate for
some representative cases. Roughly we could say that
the effect of the various noise sources is to shift,
along the $\varepsilon$ axe, the same sigmoidal shape.
What is different, between AN and DN, is that the shift
induced by the former is considerably more pronounced,
actually one order of magnitude in $\varepsilon$
when passing from $\sigma_{\rm AN}=0.03$ to $\sigma_{\rm AN}=
0.3$.

\begin{figure}
\begin{center}
%\
%\psfig{file=CS_m4-vs-noise.ps,height=2.in}
\includegraphics*[width=8cm,height=12cm,angle=270]{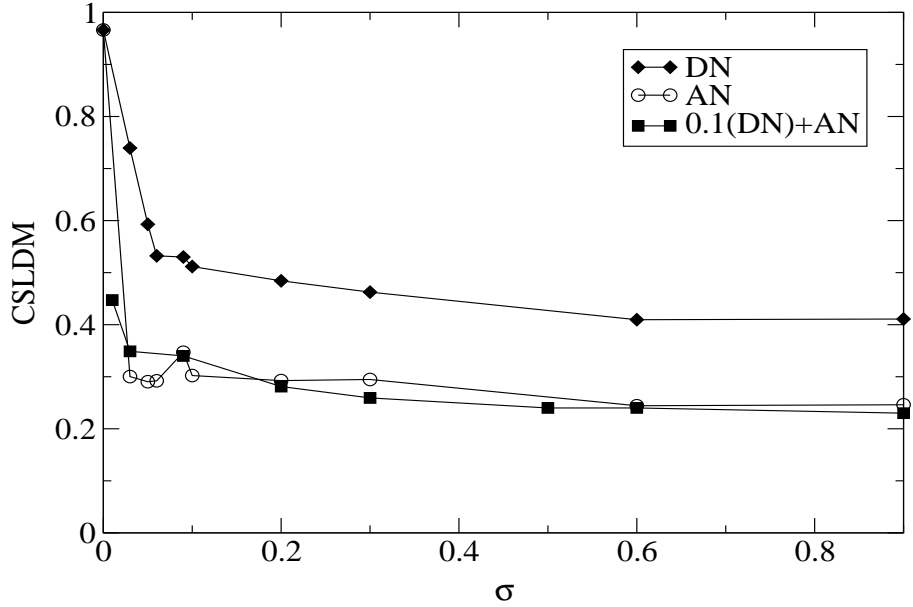}
\caption{CSLDM versus noise standard deviation, 
$\sigma$, corresponding to a four-dimensional
embedding (the other paramters being reported in the main text).
Results are shown for pure AN (no DN, open circles), pure 
DN (no AN, full diamonds) and for a system having a 
fixed level $\sigma_{\rm DN}=0.1$ plus an AN
with variable standard deviation (full squares).}
\label{CSLDMvs}
\end{center}
\end{figure}

\begin{figure}
\begin{center}
\includegraphics*[width=8cm,height=12cm,angle=270]{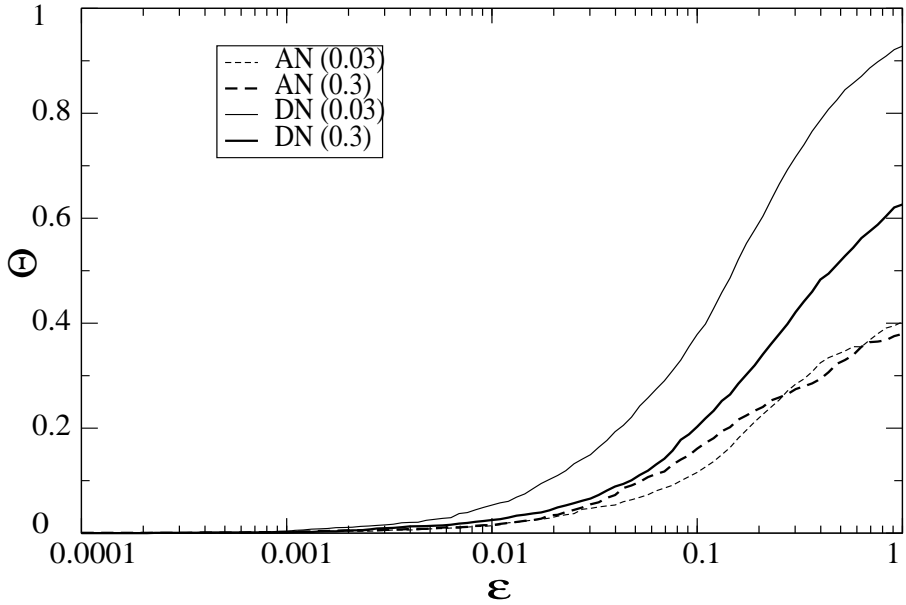}
\caption{CSLDM resolved at different $\varepsilon$ scales, for the same
levels and types of noise of Fig. \ref{CSCve}. In all cases, the measure
is estimated from a four-dimensional embedding with the
parameters indicated in the text.}
\label{CSLDMve}
\end{center}
\end{figure}

Let us now come to the analysis of the CSLDM values.
In refs. \cite{OL00,HDIM} it was argued that
this quantity is an indicator which is sensible 
to the presence of DN.
The results of fig. \ref{CSLDMvs} show that it is also
strongly affected by AN. Both the DN- and the AN-data
decrease almost monotonically with the noise level,
with the exception of a small jump at $\sigma=0.09$.
The origin of the latter is not completely clear,
though it may be partially due to the statistical
fluctuations induced by the sampling over the
$n_p$ points (eq. (\ref{e7})).
As in fig. \ref{CSCvs} we present some cases
in which the combined action of AN and DN takes place,
just to show a limitation of the present approach.
The pure AN and the AN+DN data are not
so close as in fig. \ref{CSCvs} but there is not
a definite trend to discriminate between a
non-stochastic ($\sigma_{\rm DN}=0$) and a stochastic
underlying dynamics, when the original time
series is contaminated by AN. 
As a general property we should note that the
AN-free values are always greater than
the ones in which AN is present. 
For the sake of completeness a $\Theta$-vs-$\varepsilon$
plot for the measure is given in fig. \ref{CSLDMve}.
With respect to fig. \ref{CSCve} the sigmoidal shape
is flattened (rather than shifted), and the AN curves are somehow
less regular than the DN ones. Nonetheless, the lowering
effect of noise is clearly appreciable. Note also that
in this case the effect of varying the AN level of
one order of magnitude is far less evident than in the CSC.
In fact this is consistent with fig. \ref{CSLDMvs} where
it is seen that the CSLDM jumps down and levels off to $\simeq 0.3$
as soon as a small amount of AN is introduced.

\section{Conclusions}
\label{S_Conc}
In this paper we have presented the extension of the method
of refs. \cite{OL98,OL00} to the analysis of time series affected
by AN. The basic task which one has to perform
is to compare the behaviour of the statistical continuity
of the coordinate and of the statistical differentiabilty of the
natural measure, at different levels of noise.
While the first is sensibly different from the ``clean'' one only
when AN is present, the second 
is affected by both dynamical and additive noise.
Hence, it is possible to discriminate between the two
cases. Altough we believe that the present method
can be readily applied also to high-dimensional and/or chaotic systems, the
results discussed here refer to the simple Van der Pol oscillator.
A further key step in the analysis of experimental time series,
namely the criterion to adopt when both types of noise are present,
has been only touched here and is currently investigated by
means of real physiological data.

\section{Acknowledgements}
\label{S_Ack}
This work was supported by grants of the spanish CICYT (grant no. PB96--0085),
the European TMR Network-Fractals c.n. FMRXCT980183,
the Universidad Nacional de Quilmes (Argentina) and the Universidad de
Alicante (Spain). G. Ortega is a member of CONICET Argentina.

\section{Appendix}
\label{S_App}
As pointed out in \cite{PC95} the appropriate
probability distribution for the continuity test
is the binomial one:
\begin{equation}
P(k;p,n_\delta)=\frac{n_\delta !}{k! (n_\delta-k)!} p^k 
(1-p)^{n_\delta-k} \; , \; \; k=0,1,\dots,n_\delta \; ,
\label{bin_dist}
\end{equation}
with $p=n_\varepsilon/N$, $n_\varepsilon$ 
being the number of points in $\varepsilon$-set
(see text). In addition,
due to the null hypotesis under consideration, 
the probility $p_j$ appearing in eq. (\ref{e8})
is just $p_j=p^{n_\delta}$. Except for
the trivial case $p=1$, for which one has
$P(k)=\delta_{k,n_\delta}$,
the maximum of the distribution (\ref{bin_dist})
is located at $k_{\rm max}=[p(n_\delta+1)]$
($[ \; \cdot \; ]$ denoting the integer part 
- see, for istance, \cite{MGB74}).
Now, let us observe that in the present 
case we must calculate only the ratio:
\begin{equation}
\frac{p^{n_\delta}}{P_{\rm max}}=
\frac{k_{\rm max}! (n_\delta-k_{\rm max})!}
{n_\delta !} \rho^{n_\delta-m_{\rm max}} \; ,
\label{expr_rppmax}
\end{equation}
where $\rho=p/(1-p)$.
>From a numerical point of view it is advantageous to
exploit the following trick in the left-hand side
of eq. (\ref{expr_rppmax}). Take $M=
\max{(k_{\rm max},n_\delta-k_{\rm max})}$,
so that $M!$ can be simplified
in the numerator and in the denominator of
the fraction leaving:
\begin{equation}
\frac{k_{\rm max}! (n_\delta-k_{\rm max})!}
{n_\delta !}=\frac{(n_\delta-M)!}
{n_\delta (n_\delta-1) \dots (M+1)}=
\prod_{\ell=1}^{n_\delta-M} \frac{\ell}{\ell+M} \; .
\label{trick_icb}
\end{equation}
Dealing with the product in eq. (\ref{trick_icb})
has the advantage of avoiding ratios of very large
numbers. However, the powers of $\rho$ appearing
in eq. (\ref{expr_rppmax}) can still introduce
rather small numbers, which have to be treated
through their logarithms.

\begin{thebibliography}{00}
\bibitem{WG94} A.S. Weigend and N.A. Gershenfeld (editors), {\it Time series Prediction}, Santa Fe Institute Studies in the Sciences of Complexity series, vol. XV (Addison Wesley, Reading, 1994). 
\bibitem{KS97} H. Kantz and T. Schreiber, {\it Nonlinear Time Series Analysis}
(Cambridge University Press, Cambridge, 1997).
\bibitem{KANTZ} H. Kantz, pp. 475-490 in \cite{WG94}.
\bibitem{SY99} T. D. Sauer and J. A. Yorke, Phys. Rev. Lett. {\bf 83}, 1331 (1999). 
\bibitem{SM90} G. Sugihara and R. May, Nature (London) {\bf 344}, 734 (1990).
\bibitem{KG92} D.T. Kaplan and L. Glass, Phys. Rev. Lett. {\bf 68}, 427 (1992);
Physica D {\bf 64}, 431 (1993).
\bibitem{WB93} R. Wayland, D. Bromley, D. Pickett and A. Passamante,
Phys. Rev. Lett. {\bf 70}, 580 (1993).
\bibitem{SC94} L.W. Salvino and R. Cawley, Phys. Rev. Lett. {\bf 73}, 1091 (1994).
\bibitem{BK99} J. Bhattacharya and P. P. Kanjil, Physica D {\bf 132}, 100 (1999).
\bibitem{OL98} G. Ortega and E. Louis, Phys. Rev. Lett. {\bf 81}, 4345 (1998).
\bibitem{OL00} G. Ortega and E. Louis, Phys. Rev. E. {\bf 62}, 3419 (2000).
\bibitem{TSM} M. San Miguel and R. Toral, in 
{\it Instabilities and Nonequilibrium Structures}, Vol. VI, 
E. Tirapegui and W. Zeller eds. (Kluwer Academic Pub., 1997).
\bibitem{PC95} L. Pecora, T. Carroll and J. Heagy, Phys. Rev. E. {\bf 52}, 3420 (1995).
\bibitem{SY97} T. Sauer y  J. Yorke, Ergodic Th. Dyn. Syst. {\bf 17}, 941 (1997).
\bibitem{Mi96} T. Miyano, Int. J. Bif. Chaos, {\bf 6}, 2031 (1996) 
\bibitem{BP96} M. Barahona and C.-S. Poon, Nature {\bf 381}, 215 (1996).
\bibitem{HDIM}  G. Ortega, C. Degli Esposti Boschi and E. Louis,
{\it ``Detecting determinism in high-dimensional chaotic systems''},
submitted to Phys. Rev. E.
\bibitem{NOTE} Note that in eq. (\ref{ixj}) we have assumed
that the $G_{jk}$ are independent of $\vec{x}$.
However, the generalisation to the $\vec{x}$-dependent
case does not modify our picture since
the essential point for the continuity property is
the existence of a ${(\delta t)}^{1/2}$ term.
This turn out to be the case when the increments
are calculated through the Milshtein scheme \cite{TSM}.
\bibitem{Sz93} G. G. Szpiro, Physica D {\bf 65}, 289 (1993).
\bibitem{MGB74} A. M. Mood, F. A. Graybill and D. C. Boes,
{\it Introduction to the theory of statistics} (McGraw Hill, Inc., 1974).
\bibitem{AB93} H. Abarbanel, R. Brown, J. Sidorowich and L. Tsimring,
Rev. Mod. Phys. {\bf 65}, 1331 (1993).
\bibitem{CF00} M. Cencini, M. Falcioni, E. Olbrich, H. Kantz and A. Vulpiani,
Phys. Rev. E {\bf 62}, 427 (2000).
\bibitem{JK99} J. Jeong, M.S. Kim and S.Y. Kim,
Phys. Rev. E {\bf 60}, 831 (1999).
\bibitem{DG93} M. Ding {\it et al.} Physica D {\bf 69}, 404 (1993).
\bibitem{GM98} A. Galka, T. Maab and G. Pfister,
Physica D, {\bf 121}, 237 (1998).
\bibitem{CE91} M. Casdagli, S. Eubank, D. Farmer and J. Gibson,
Physica D {\bf 51}, 52 (1991).
\bibitem{HB98} R. Hegger, M. J. B\"{u}nner and H. Kantz, Phys. Rev. Lett. {\bf 81(3)}, 558 (1998).
\bibitem{OK97} E. Olbrich and H. Kantz, Phys. Lett. A {\bf 232(1-2)}, 63 (1997).
\bibitem{TE92} J. Theiler, S. Eubank, A. Longtin, B.
Galdrikian and J. D. Farmer, Physica D {\bf 58}, 77 (1992).
\bibitem{OP89} A. Osborne and A. Provenzale, Physica D {\bf 35}, 357 (1989).
\bibitem{Sc98} T. Schreiber, Phys. Rev. Lett {\bf 80}, 2105 (1998).
\end{thebibliography}
\end{document}